\documentclass[aps,prL,twocolumn,longbibliography,groupedaddress,showpacs,floatfix,superscriptaddress]{revtex4-2}
\usepackage{graphicx}  
\usepackage{dcolumn}   
\usepackage[english]{babel}
\usepackage{bm}       
\usepackage{amsmath, amsthm, amssymb}
\usepackage{bbold}
\usepackage{amsmath}
\usepackage{mathrsfs}
\usepackage{array}
\usepackage{multirow}
\usepackage{physics}
\usepackage[dvipsnames,usenames]{color}
\usepackage{soul} 
\usepackage{ulem} 
\usepackage{orcidlink}

\usepackage{hyperref}
\hypersetup{
    citecolor=magenta,
    colorlinks=true,
    linkcolor=blue,
    filecolor=magenta,      
    urlcolor=blue,
    pdftitle={Overleaf Example},
    pdfpagemode=FullScreen,
    }

\emergencystretch=\maxdimen
\newcommand{\calH}{\mathcal{H}}

\begin{document}

\normalem

\title{Quantum percolation in honeycomb lattices under random spin–orbit coupling}

\author{W. S. Oliveira \orcidlink{0000-0001-8892-845X}}
\affiliation{Departamento de Física, PUC-Rio, 22452-970, Rio de Janeiro RJ, Brazil}
\author{Julián Faúndez \orcidlink{0000-0002-6909-0417}}
\affiliation{Departamento de Física y Astronomía, Universidad Andres Bello, Santiago 837-0136, Chile}
\author{Welles Morgado  \orcidlink{0000-0002-9690-6258}}
\affiliation{Departamento de Física, PUC-Rio, 22452-970, Rio de Janeiro RJ, Brazil}

\begin{abstract}
We investigate quantum percolation in a honeycomb lattice with site dilution and random spin--orbit coupling. Using exact diagonalization combined with finite-size scaling analysis, we study the metal-insulator transition, extracting the quantum percolation threshold $p_q$, and the correlation-length exponent, $\nu$. In the absence of spin–orbit coupling, we find that $p_q$ remains finite and demonstrate that the quantum threshold is significantly higher than the classical site-percolation threshold $p_c$ of the honeycomb lattice. When spin--orbit coupling is present, the spectral statistics exhibit a crossover from the Gaussian orthogonal ensemble to the Gaussian symplectic ensemble, reflecting the change in symmetry class. Simultaneously, the quantum percolation threshold shifts systematically to lower occupation probabilities, indicating that the spin--orbit coupling favors delocalization. For sufficiently strong spin--orbit coupling, $p_q$ tends to saturate, while the critical exponent approaches the expected one of the two-dimensional symplectic universality class.
\end{abstract}

\date{\today}

\pacs{
64.60.Ak 
64.60.Fr 
71.55.Jv 
72.15.Rn 
}
\maketitle


\section{Introduction}

Understanding how disorder, quantum interference, and symmetry shape the electronic transport in low-dimensional systems remains a central challenge in condensed matter physics.
Graphene and related two-dimensional honeycomb materials constitute a paradigmatic setting for addressing this problem, owing to their Dirac-like spectrum -- characterized by linearly dispersing states around the Dirac points --and the resulting unconventional quantum transport properties in the clean limit \cite{Hatsugai2006,Nomura2007,Fritz2011,CastroNeto2009,Novoselov2005,Peres2010}. A central issue in these systems is the effect of disorder on electronic conduction, particularly whether impurities, vacancies, or lattice dilution can drive a disorder-induced metal--insulator transition (MIT) in two dimensions \cite{Peres2010,Mucciolo2010,Evers2008}. This problem becomes even richer in the presence of spin-orbit coupling (SOC), which, although weak in pristine graphene, can be strongly enhanced by curvature, adatoms, substrate proximity, or chemical functionalization \cite{KaneMele2005,HuertasHernando2006,Weeks2011,Avsar2014,Wang2015}. Depending on its form and strength, SOC can modify the band structure, change the symmetry class, affect localization properties, and even stabilize topologically nontrivial phases \cite{KaneMele2005,GmitraFabian2015}. For this reason, these systems provide a particularly suitable framework for investigating the interplay between quantum interference, localization, and disorder-driven MIT in two dimensions, phenomena that have been experimentally detected in graphene-based heterostructures where proximity-induced SOC produces distinct features in magnetotransport measurements \cite{Wang2015,Avsar2014}.

A standard theoretical framework for addressing disorder-driven transport is the  Anderson Impurity Model (AIM), in which electrons propagate on a lattice with random on-site energies \cite{Anderson58,Kostadinova2019}. This \textit{diagonal disorder} models fluctuations in the local potential landscape and has played a central role in the theory of localization and MIT. In two dimensions, the Anderson problem is especially subtle because quantum interference strongly suppresses diffusion, so that the transport properties depend crucially on the symmetry class of the Hamiltonian \cite{Abrahams79,Asada2002}. In honeycomb lattices, the Dirac-like spectrum and the possibility of intervalley scattering already make the localization problem nontrivial, while SOC enriches it further by breaking spin-rotation symmetry and driving the system into the symplectic universality class \cite{Evangelou1995,Asada2002}. In this case, unlike the conventional spinless two-dimensional problem, a disorder-driven MIT may occur \cite{Asada2002}. For honeycomb structures, this interplay between on-site disorder and SOC has been directly investigated in tight-binding approaches, where increasing diagonal disorder was shown to induce a MIT with critical behavior consistent with the symplectic class \cite{Queiroz2015}. 



A closely related, but conceptually distinct, scenario arises when hopping between neighboring sites is suppressed by missing sites, vacancies, or defect-induced broken bonds. This form of \emph{off-diagonal disorder} is analogous to the classical percolation problem \cite{stauffer18}, in which the global connectivity of the lattice is controlled by the occupation probability $p$ of active sites or bonds. In the electronic problem, however, the existence of a spanning cluster does not by itself ensure transport, since quantum interference may still localize the particle. For this reason, the system is described by the quantum percolation model (QPM). In this context, $p_c$ denotes the classical percolation threshold, below which no spanning cluster exists, whereas $p_q$ marks the threshold for quantum delocalization; typically, $p_c < p_q$ due to interference effects~\cite{Oliveira21}. For honeycomb systems, this distinction is particularly relevant because the vacancy limit can be viewed as a quantum site-percolation problem \cite{Schubert2012}. Despite extensive studies, no precise value for the quantum site-percolation threshold in two-dimensional honeycomb lattices exists; some studies offer only a rough estimate of its magnitude range, merely indicating that the threshold significantly exceeds the classical value and is strongly affected by finite-size effects \cite{Chandrashekar2013,Feng2023,Kostadinova2019}, whereas its dependence on additional physical ingredients such as spin-orbit coupling remains largely unexplored. This makes it especially interesting to determine the precise value of $p_q$, how additional ingredients, such as SOC, modify its value, and the respective critical exponents. At this point, it's worth emphasizing that, although both the AIM and the QPM exhibit localization phenomena, the nature of the disorder threshold in the latter was historically unsettled. Some early approaches supported complete localization for any finite disorder \cite{Avishai92,Soukoulis91,Shapir82,Raghavan81,Raghavan84}, whereas others pointed to a finite critical threshold above which extended states could emerge \cite{Nazareno02,Islam08,Schubert08,Gong09,Dillon14,Yu05,Oliveira25}. 


In this work, we provide a systematic study of quantum site-percolation in honeycomb lattices in the presence of random SOC. Using a tight-binding formulation combined with exact diagonalization and finite size scaling (FSS), we determine the quantum percolation threshold $p_{q}$ and the associated critical exponent $\nu$, and analyze how these quantities evolve as a function of SOC strength.
%
%
The remainder of this paper is organized as follows. In Sec.~\ref{model}, we introduce the model and methods; in Sec.~\ref{results}, we present and discuss the results; and in Sec.~\ref{conclusions}, we summarize our main findings.


\section{Model and Method}
\label{model}
\subsection{Hamiltonian}
We study non-interacting spinful electrons on a two-dimensional honeycomb lattice with periodic boundary conditions (PBC) imposed in both spatial directions. The lattice contains $L\times L$ unit cells, each comprising two sublattice sites. Consequently, the total number of lattice sites is $N_{\mathrm{sites}} = 2L^2$, and when spin is taken into account, the single-particle Hilbert space has dimension $\mathcal{N} = 4L^2$.
The complete Hamiltonian with spin-orbit coupling is defined as
\begin{equation}
\mathcal{H}
=
\sum_{\langle ij\rangle}
\eta_i \eta_j
\left[
a_{i}^\dagger
\left(
\mathbf{1}_2
+
i\mu\,\mathbf{V}_{ij}\cdot\boldsymbol{\sigma}
\right)
a_{j}
+
    \mathrm{H.c.}
\right],
\label{eq:Hamiltonian}
\end{equation}
where $\langle ij\rangle$ labels nearest-neighbor pairs on the honeycomb lattice, and $\boldsymbol{\sigma}=(\sigma_x,\sigma_y,\sigma_z)$ represents the Pauli matrices operating in spin space. The parameter $\mu$ sets the magnitude of the spin-dependent hopping amplitude.
Quantum site percolation is implemented through binary occupation variables $\eta_i \in \{0,1\}$ drawn independently for each lattice site according to $\mathbb{P}(\eta_i = 1) = p$, and $\mathbb{P}(\eta_i = 0) = 1-p$, where $p$ is the occupation probability. The factor $\eta_i\eta_j$ in Eq. (\ref{eq:Hamiltonian}) guarantees that hopping processes take place only when both sites are occupied.

The spin-dependent matrix associated with each nearest-neighbor pair is given by
$ T_{ij} = \mathbf{1}_2 + i\mu\,\mathbf{V}_{ij}\cdot\boldsymbol{\sigma}$,
where $\mathbf{V}_{ij}=(V^x_{ij},V^y_{ij},V^z_{ij})$ is a random, three-component vector \cite{Queiroz2015,Evangelou1995}. Each component is independently drawn from a uniform distribution, $V^\alpha_{ij} \in [-1/2,\,1/2]$.
Writing out the term $T_{ij}$ explicitly, we find
\begin{equation}
T_{ij}
=
\begin{pmatrix}
1+i\mu V^z_{ij}
&
\mu V^y_{ij}+i\mu V^x_{ij}
\\
-\mu V^y_{ij}+i\mu V^x_{ij}
&
1-i\mu V^z_{ij}
\end{pmatrix}.
\label{eq2}
\end{equation}
The term proportional to $\mu$ introduces spin-dependent hopping amplitudes, generating both spin-conserving and spin-flip processes. Hermiticity of the Hamiltonian is ensured by inclusion of the Hermitian conjugate.  The choice of random SOC is motivated by the fact that it allows us to probe the combined effect of disorder and symmetry on quantum percolation. In contrast to a uniform SOC, which acts as a coherent modification of the hopping amplitudes, random SOC introduces dependent spin scattering and therefore has a more direct impact on localization and criticality \cite{Queiroz2015,Evangelou1995}.

\subsection{Symmetry of  the Hamiltonian}
\label{subsec:symmetry}
To clarify the microscopic mechanism behind the symmetry-class crossover and its physical implications, we examine how spin-rotation symmetry, time-reversal invariance, and SOC are present in our model. In the absence of SOC ($\mu/t=0$), the Hamiltonian [Eq. (\ref{eq:Hamiltonian})] describes a spin-degenerate tight-binding model with site dilution. In this regime, the system maintains complete SU(2) spin-rotation symmetry, so spin remains a conserved quantum number. Time-reversal symmetry (TRS) is also preserved and is implemented by the antiunitary operator $\mathcal{T}=i\sigma_{y}\mathcal{K}$, which obeys $\mathcal{T}^{2}=-1$ for spin-1/2 particles. However, due to the SU(2) symmetry, the Hamiltonian can be split into two decoupled spin sectors, effectively mapping the problem onto two independent replicas of a spinless system. As a result, the spectral correlations—understood as the statistical correlations between neighboring energy levels—obey the Gaussian Orthogonal Ensemble (GOE), and quantum interference gives rise to pronounced localization effects in two dimensions \cite{Evers2008,Atas2013}. 

The presence of SOC ($\mu\neq 0$) qualitatively changes the situation. The SOC-dependent factor [Eq. (\ref{eq2})] generates random spin-dependent hopping amplitudes that entangle spin and orbital degrees of freedom. As a consequence, SU(2) spin-rotation symmetry is explicitly broken, spin is no longer conserved, and spin precession is induced along electronic paths \cite{Evangelou1995}. Crucially, because the SOC under consideration is nonmagnetic, time-reversal symmetry is still preserved, and the relation $\mathcal{T}^{2}=-1$ remains valid \cite{KaneMele2005}. The combination of broken spin-rotation symmetry and preserved TRS places  the system in the  Gaussian Symplectic Ensemble (GSE) \cite{Evers2008,Asada2002}. A direct implication of $\mathcal{T}^{2}=-1$ is Kramers degeneracy, in which every eigenstate has a degenerate counterpart related by time reversal. This degeneracy reduces backscattering and diminishes localization via quantum interference, leading to weak anti-localization effects \cite{Hikami1980}. As a result, the extended states become more robust to disorder. The crossover between symmetry classes is directly captured by the spectral statistics through $\langle r \rangle$, which evolves from the GOE value toward the GSE prediction as the SOC strength increases. This behavior demonstrates that the system undergoes a symmetry-driven transition from GOE $\leftrightarrow$ GSE. \cite{Atas2013,Asada2002}.


\subsection{The ratio of adjacent gaps}
The crossover between localized and delocalized regimes can be primarily interpreted as a modification in how broadly the wave function extends over the tight-binding orbitals on a lattice. Consequently, in the Hilbert space, the localized phase is associated with non-ergodic eigenstates, and its spectrum does not exhibit fractal properties \cite{Nicolas2019}. In contrast, the delocalized phase features ergodic eigenstates. The critical point (or critical phase, when present) is expected to show multifractal characteristics \cite{Nicolas2019}. To analyze the metal–insulator transition in our model, we use spectral correlations, a standard tool for discriminating between extended and localized phases via their universal signatures, which depend only on the symmetry class~\cite{Evers08,luo2021universality}.
Let $\epsilon_{\alpha}$ be the eigenvalues of $\calH$. Defining the level spacings $s_{\alpha} = \epsilon_{\alpha+1} - \epsilon_{\alpha}$, we compute the \emph{ratio of adjacent gaps}~\cite{Atas13},
\begin{equation}
\label{ratio_gaps}
r_{\alpha} \;=\; \frac{\min(s_{\alpha}, s_{\alpha+1})}
                     {\max(s_{\alpha}, s_{\alpha+1})},
\end{equation}
and obtain the mean ratio $\langle r \rangle$ by averaging $r_{\alpha}$ over $\alpha$. In the delocalized regime with intact time-reversal symmetry, the statistics of $r$ follow the Wigner–Dyson (WD) distribution corresponding to the GOE of random matrix theory, giving $\langle r \rangle \approx 0.53$~\cite{Atas13}. In the localized regime, on the other hand, energy levels become effectively uncorrelated and obey Poisson statistics, leading to $\langle r \rangle \approx 0.39$~\cite{Atas13}. Introducing SOC, the symmetry class changes from GOE to GSE, while time-reversal symmetry remains intact (with $T^{2}=-1$ due to Kramers degeneracy); consequently, for this case, in the delocalized regime the WD statistics are described by the GSE, yielding $\langle r \rangle \approx 0.68$~\cite{Atas13}.

\subsection{The participation entropy}

To characterize the structure and localization properties of the eigenstates, we consider the participation entropy (PE), defined as the Rényi entropy in Hilbert space~\cite{Liu25,Sierant22}. For a normalized state \(\ket{\Psi}=\sum_j c_j\ket{j}\) in the basis \(\{\ket{j}\}\), the PE of order \(q\) reads~\cite{Luitz2014}
\begin{equation}
S_q=\frac{1}{1-q}\ln I_q,
\end{equation}
where
\begin{equation}
I_q=\sum_j |c_j|^{2q}
\end{equation}
is the generalized inverse participation ratio. In this work, we focus on the case \(q=2\), for which
\begin{equation}
S_2=-\ln I_2,
\qquad
I_2=\sum_j |c_j|^4.
\end{equation}
The generalized IPR scales with Hilbert-space dimension \(\mathcal N\) as
\begin{equation}
I_q \sim \mathcal{N}^{-(q-1)D_q},
\end{equation}
so that for \(q=2\),
\begin{equation}
I_2 \sim \mathcal{N}^{-D_2}.
\label{eq:IPR_q2}
\end{equation}
Here \(\mathcal N=4L^2\), and \(D_2\) is extracted from finite-size scaling.
In the localized phase, the electron occupies only a finite number of lattice sites, so $I_2$ is essentially independent of $L$; Eq.~\eqref{eq:IPR_q2} thus implies $D_2=0$ and $S_2\sim \text{const}$. In the extended phase, amplitudes scale as $|c_j|^2\sim 1/\mathcal{N}$, yielding $I_2\sim 1/\mathcal{N}$, hence $D_2=1$ and $S_2\sim \ln \mathcal{N}$. At the critical point (or in a critical phase), one expects multifractal behavior with $0<D_2<1$~\cite{Ujfalusi14}; correspondingly,
\begin{equation}
S_2 \sim D_2 \ln \mathcal{N}.
\end{equation}
Therefore, $S_2/\ln \mathcal{N}$ provides an estimate of $D_2$, smoothly interpolating between $0$ (fully localized) and $1$ (fully extended).

\section{Results}
\label{results}

Throughout this paper, we focus on the energy window \(E/t = 0.6 \pm 0.2\). This choice does not imply a qualitative loss of generality. Previous studies have established that changing the energy window mainly leads to quantitative shifts in the strength and location of the quantum-percolation mobility edge, reflecting the underlying energy dependence of the density of states \cite{Gong09,Dillon14,Oliveira25,Islam08}. In contrast, the scaling itself is expected to be robust across different energy windows. For each lattice size \(L\) and occupation probability \(p\), disorder averaging is performed over \(10^{6}/\mathcal{N}\) independent realizations, ensuring comparable statistical accuracy across system sizes.

\begin{figure}[t]
    \centering
    \includegraphics[width=0.9\linewidth]{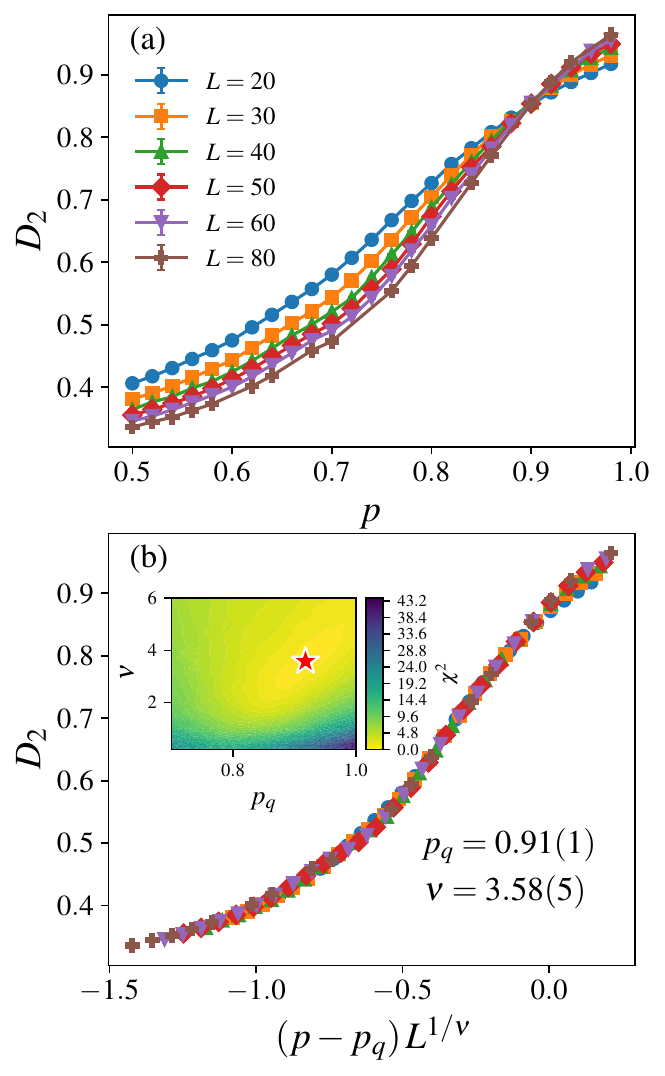}
    \caption{ (a) Approximant of the fractal dimension $D_{2}$ as a function of the site occupation probability $p$, for several lattice sizes $L$. (b) Data collapse of $D_{2}$ in the vicinity of the critical $p_{c}$. Inset: colormap of the correlated critical exponent $\nu$ and $p_q$ in the $(\nu, p)$ plane. The minimum of $\chi^{2}$ (marked by the red star) identifies the optimal estimates of $\nu$ and $p_q$. When not shown, error bars are smaller than symbol sizes.} 
    \label{fig:collapse_mu00}
\end{figure}

\subsection{Absence of spin-orbit coupling: $\mu/t = 0$}

First, we study the quantum site-percolation problem in the absence of SOC, since the corresponding estimates of $p_q$ for the standard case have not been previously investigated. In this regime, the Hamiltonian simplifies to a spin-degenerate nearest-neighbor tight-binding model with site dilution, which falls into the orthogonal symmetry class \cite{Atas13}. Fig. \ref{fig:collapse_mu00}\,(a) shows the approximant of fractal dimension $D_{2}$, obtained from the participation entropy, as a function of the occupation probability $p$, for several lattice sizes $L$. The quantity $D_{2}$ serves as a direct measure of the spatial structure of the eigenstates. For small $p$, $D_{2}$ decreases as the lattice size increases, indicating localization. As $p$ grows, the curves bend upward and gradually approach $D_{2} \approx 1$, indicating an extended regime. The crossing near $p \approx 0.9$ signals the quantum percolation threshold $p_q$, where the system becomes scale-invariant. To extract the critical parameters quantitatively, we assume a one-parameter finite-size scaling form \cite{Fisher71,Barber83},
$
D_{2}(p,L)=f\!\left[(p-p_q)L^{1/\nu}\right],
$
where $\nu$ is the correlation-length exponent. When $D_2$ is plotted as a function of the scaling variable $(p-p_q)L^{1/\nu}$, data for different $L$ and $p$ should collapse onto a single universal curve, as shown in Fig.~\ref{fig:collapse_mu00}(b). A least-squares fit yields the optimal estimate of $p_q$ and $\nu$ (see inset of Fig.~\ref{fig:collapse_mu00}(b)), giving $p_q = 0.91(1)$ and $\nu = 3.58(5)$. As expected, the quantum percolation threshold $p_q$ exceeds the classical percolation threshold of this lattice, $p_c = 0.6970(1)$ \cite{Feng2008}. The extracted critical exponent $\nu$ is different from estimations of the two-dimensional symplectic universality class ($\nu \approx 2.7$) \cite{Queiroz2015,Asada2002,Asada2004}, but still satisfies the inequality $\nu \geq 2/d$ \cite{Chayes86}.

\begin{figure}[t]
    \centering
    \includegraphics[width=1.0\linewidth]{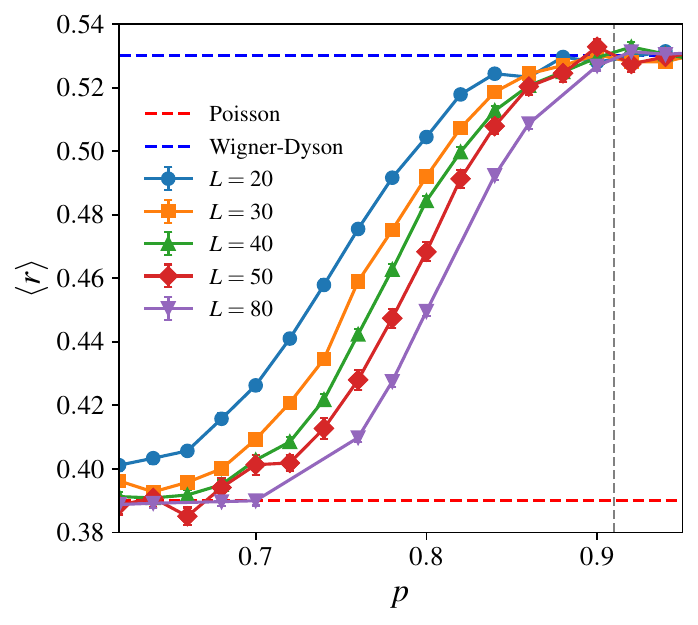}
    \caption{The average ratio of adjacent gaps $\langle r \rangle$ as a function of the site occupation probability $p$ for different lattice sizes $L$. The blue and red dashed lines correspond to Poisson and Wigner–Dyson statistics, respectively.} 
    \label{fig:gap_mu00}
\end{figure}

Independent evidence for the transition comes from the average ratio of adjacent gaps, $\langle r \rangle$, plotted in Fig. \ref{fig:gap_mu00}. In the localized phase, $\langle r \rangle$ follows a Poisson distribution (dashed red line) with $\langle r \rangle\approx 0.39$, while in the extended phase it obeys a Wigner–Dyson distribution (dashed blue line) with $\langle r \rangle\approx 0.53$. As $p$ increases, $\langle r \rangle$ varies smoothly between these two limiting values, and the curves for different system sizes 
saturates at the $p_q = 0.91(1)$.

\subsection{Presence of spin-orbit coupling: $\mu/t \ne 0$}

\begin{figure}[t]
    \centering
    \includegraphics[width=1.0\linewidth]{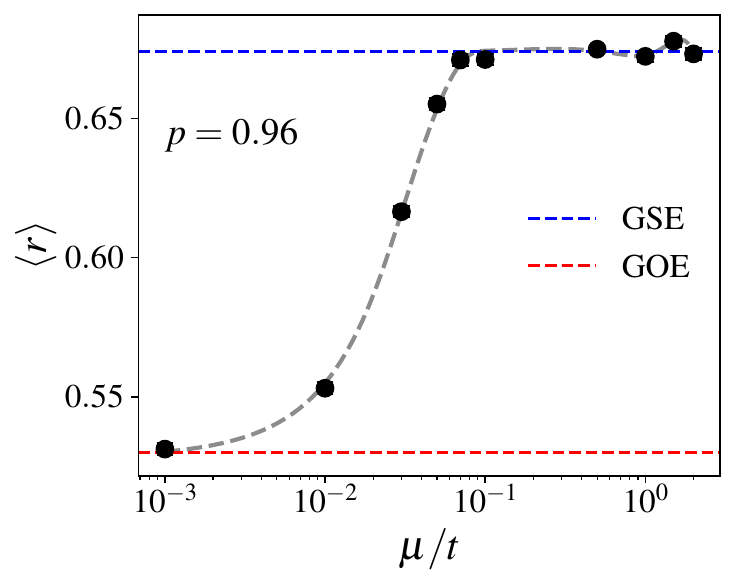}
    \caption{Average ratio of adjacent gaps $\langle r \rangle$, as a function of the spin-orbit coupling (SOC) strength $\mu/t$ for an occupation probability $p=0.96$. The horizontal dashed lines indicate the universal random-matrix predictions for the Gaussian symplectic ensemble (blue line) and the Gaussian orthogonal ensemble (red line), respectively. In addition, the dashed gray line is a guide to the eye.} 
    \label{fig:mu_vs_r}
\end{figure}

After discussing the case without SOC, we now turn to the regime $\mu /t\neq 0$. The first point to address is the change in the symmetry class of the Hamiltonian induced by SOC. In the absence of SOC, and for a disordered system with time-reversal symmetry and SU(2) spin-rotation invariance, the level statistics are expected to belong to the GOE. Once SOC is introduced, the SU(2) spin-rotation symmetry is broken. However, for nonmagnetic SOC in a spin-$\tfrac12$ system, time-reversal symmetry is still preserved, with $T^2=-1$, as previously discussed.
\begin{figure}[b]
    \centering
    \includegraphics[width=1.0\linewidth]{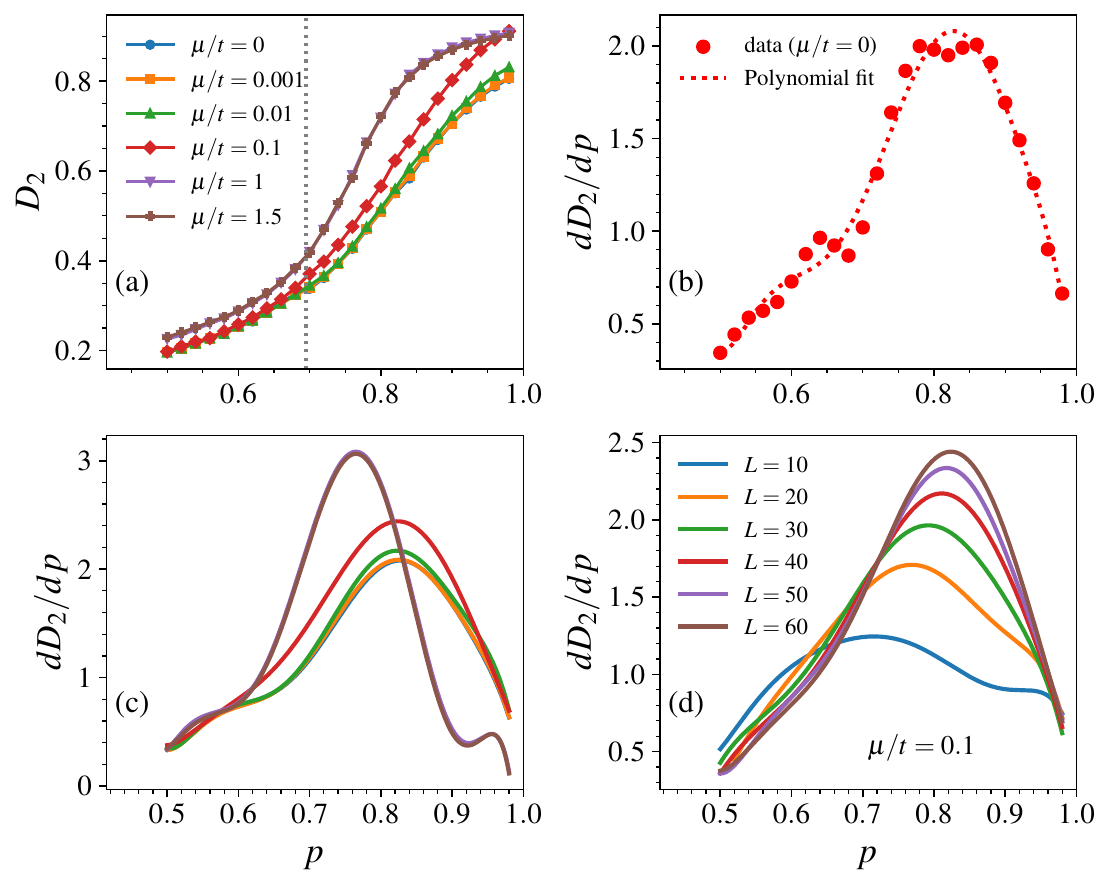}
    \caption{(a) Approximant of the fractal $D_2$ as a function of the occupation probability $p$, shown for several values of spin-orbit coupling (SOC) strength $\mu/t$. The dotted line signals the classical percolation threshold, $p_c$. (b) Derivative of the participation entropy $D_2$ for the case $\mu/t=0$: the symbols represent the numerical data, while the dash-dotted line is a polynomial fit used to capture the overall trend. (c) Derivative $dD_2/dp$ obtained from polynomial fits to each $D_2(p)$ curve, for the same set of $\mu/t$ values shown in panel (a). Same as (c) but for fixed $\mu/t = 0.1$ and different lattice sizes $L$.} 
    \label{fig:D2_derivative}
\end{figure}
\begin{figure*}
    \centering
    \includegraphics[width=1.0\linewidth]{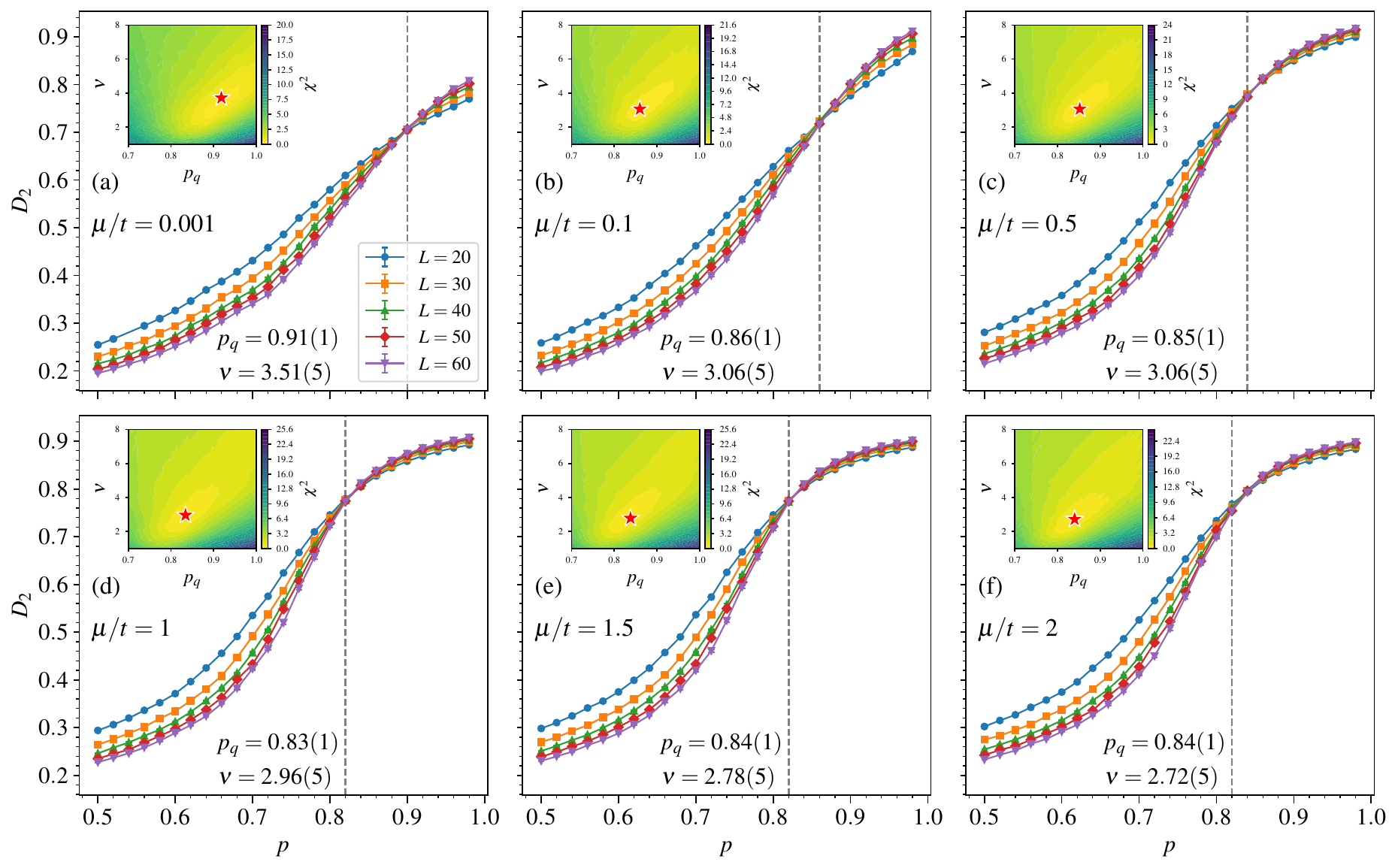}
    \caption{Approximant of the fractal dimension $D_{2}$ as a function of the site occupation probability $p$ for various lattice sizes $L$ and spin–orbit coupling (SOC) parameter $\mu/t$. The dashed vertical lines indicate the critical value $p_{q}$. The insets show the optimal $p_q$ and $\nu$ extracted from the  finite size scaling (FSS) analysis. When not shown, error bars are smaller than the symbol sizes.}
    \label{fig:entropy_vs_mu}
\end{figure*}
As a consequence, the honeycomb lattice is no longer in the orthogonal class, but instead crosses over to the symplectic universality class, whose spectral correlations are described by the Gaussian symplectic ensemble (GSE).

This change in symmetry class can be observed through $\langle r\rangle$, whose universal random-matrix values are $\langle r\rangle_{\mathrm{GOE}} \approx 0.5307$ for the GOE and $\langle r\rangle_{\mathrm{GSE}} \approx 0.6744$ for the GSE \cite{Atas13}. In Fig.~\ref{fig:mu_vs_r}, we plot $\langle r\rangle$ as a function of the SOC strength $\mu/t$ for a fixed $p=0.96$, which lies inside the delocalized region. The purpose of this analysis is to determine how the increase of SOC modifies the universality class of the level statistics. For very small values of $\mu/t$, the effect of SOC on the spectrum is still weak, and the value of $\langle r\rangle$ remains close to the GOE prediction. This indicates that, in this regime, the spectral correlations are still effectively governed by the GOE class. As $\mu/t$ increases, however, $\langle r\rangle$ changes smoothly from the GOE value and approaches the GSE prediction. From the data, this crossover becomes clearly visible around $\mu/t \approx 0.1$, above which the level statistics are already very close to the GSE limit.

To see whether this trend is also present in the eigenstate structure, we also examine the behavior of $D_2$ when SOC is present. Figure \ref{fig:D2_derivative} (a) shows $D_2$ versus the occupation probability $p$ for several values of $\mu/t$ and for fixed $L=60$. In all cases, SOC enhances the overall value of $D_2$, indicating a tendency to delocalization. This effect is relatively weak below the classical percolation threshold $p_c$, but becomes more pronounced for $p>p_c$. As we can see, the evolution of $D_2$ alone does not provide a precise estimate of the transition. A more sensitive quantity is the derivative $dD_2/dp$, whose maximum gives a qualitative estimate of the metal-insulator crossover \cite{Suntajs2021,Nicolas2019}. Since numerical differentiation is noisy for a limited set of data points, we first fit the $D_2(p)$ data with a polynomial function and then differentiate the fitted curve, as illustrated in Fig. \ref{fig:D2_derivative} (b). Repeating this procedure for different values of $\mu/t$ gives the results shown in Fig.~\ref{fig:D2_derivative} (c). For weak SOC, the position of the maximum in $dD_2/dp$ changes only slightly. By contrast, for $\mu/t \gtrsim 0.1$, the peak shifts significantly toward smaller values of $p$. This behavior indicates that SOC pushes the quantum threshold to lower occupation probabilities, i.e., it promotes delocalization. Still, the peak remains above $p_c$, as we should expect. Although this analysis provides useful qualitative insight into the transition, it is still affected by finite-size effects, as shown in Fig.~\ref{fig:D2_derivative} (d), for $\mu/t = 0.1$. Therefore, a finite-size scaling analysis is required to extract more precise information about the critical point.

To this end, we analyze the behavior of $D_2$, following the same strategy employed in the $\mu/t=0$ case. Figure~\ref{fig:entropy_vs_mu} shows $D_2$ as a function of $p$ for several $L$ and for different values of SOC strength $\mu/t$.
%
%
The first point to note is that for a very small SOC, $\mu/t=0.001$ [Fig.~\ref{fig:entropy_vs_mu}(a)], the estimated threshold $p_q$ remains equal, within error bars, to the value obtained in the absence of SOC, $\mu/t=0$. This behavior is fully consistent with the results of Fig.~\ref{fig:mu_vs_r}, where the corresponding value of $\langle r\rangle$ is still close to the GOE prediction, indicating that such a weak SOC has only a negligible effect on the honeycomb lattice. As the SOC is increased, however, a clear modification of the finite-size behavior of $D_2$ emerges. In particular, the curves show a stronger increase with system size in the critical region, accompanied by a systematic shift of the quantum percolation threshold toward smaller values of $p$. This indicates that SOC favors delocalization, allowing extended states to survive down to lower site-occupation probabilities. For larger values of $\mu/t$, this shift becomes progressively weaker, suggesting a saturation of the threshold around $p_q \approx 0.8$.
To determine $p_q$ and the critical exponent $\nu$ more accurately, we apply the same FSS procedure used in Fig.~\ref{fig:collapse_mu00}, based on the scaling form introduced for the $\mu=0$ case. The optimal values of $p_q$ and $\nu$ are obtained by minimizing the corresponding $\chi^2$ function, and the resulting cost-function maps are shown in the insets of Fig.~\ref{fig:entropy_vs_mu}. The extracted parameters are summarized in Table~\ref{tab:pc_nu_vs_lambda}. The results reveal a systematic reduction of the quantum percolation threshold as $\mu/t$ increases, together with a variation of the critical exponent $\nu$.
\begin{table}[b]
\centering
\renewcommand{\arraystretch}{1.25}
\setlength{\tabcolsep}{8pt}
\begin{tabular}{c|cc}
\hline
\multirow{2}{*}{$\mu/t$} & \multicolumn{2}{c}{$E/t = 0.6$} \\
\cline{2-3}
 & $p_q$ & $\nu$ \\
\hline
0.0    & 0.91(1) & 3.58(5) \\
0.001  & 0.91(1) & 3.51(5) \\
0.01   & 0.90(1) & 3.42(5) \\
0.05   & 0.87(1) & 3.25(5) \\
0.1    & 0.86(1) & 3.06(5) \\
0.5    & 0.85(1) & 3.06(5) \\
1.0    & 0.83(1) & 2.96(5) \\
1.5    & 0.84(1) & 2.78(5) \\
2.0    & 0.84(1) & 2.72(5) \\
2.5    & 0.84(1) & 2.73(5) \\
\hline
\end{tabular}
\caption{Quantum percolation threshold $p_q$ and correlation-length critical exponent $\nu$ as a function of the spin-orbit coupling (SOC) parameter $\mu/t$, for fixed energy $E/t = 0.6$.}
\label{tab:pc_nu_vs_lambda}
\end{table}

\begin{figure}[t]
    \centering
    \includegraphics[width=1.0\linewidth]{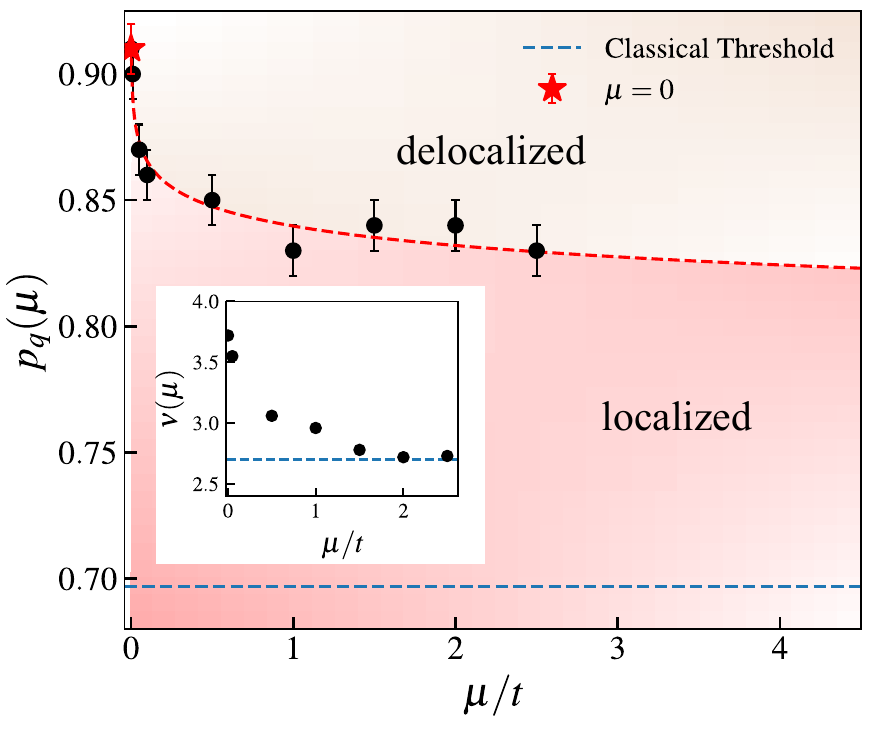}
    \caption{Phase diagram of the quantum percolation threshold $p_q$,  versus the spin–orbit coupling (SOC) strength, $\mu/t$. The red star marks the value of $p_q$ in the absence of SOC. The red dashed line is a guide to the eye, while the blue dashed line denotes the classical percolation threshold. The inset shows the dependence of the correlation-length critical exponent $\nu$ versus $\mu/t$. The horizontal dashed blue line marks the corresponding value of the symplectic case, $\nu \approx 2.7$.}
    \label{fig:diagram}
\end{figure}

The final phase diagram is presented in Fig.~\ref{fig:diagram}, where the dependence of the quantum percolation threshold $p_q$ on the SOC strength $\mu/t$ can be seen clearly. The most evident feature is the monotonic decrease of $p_q$ as $\mu/t$ increases, followed by a tendency toward saturation at larger SOC. For comparison, the horizontal blue dashed line indicates the classical percolation threshold of the honeycomb lattice, $p_c \approx 0.697$. Throughout the whole range of $\mu/t$ considered here, we find $p_q > p_c$, showing that the onset of quantum transport still requires a larger occupation probability than the purely geometrical connectivity threshold. In other words, even in the presence of strong SOC, quantum interference effects remain sufficiently important to keep the quantum percolation threshold above the classical one. It is also instructive to follow the behavior of the critical exponent $\nu$ as the SOC strength increases. This is shown in the inset of Fig.~\ref{fig:diagram}. We observe a trend similar to that found for $p_q$, as $\nu$ decreases as $\mu/t$ increases and then tends to saturate for larger values of SOC. Interestingly, the saturation occurs near $\nu \approx 2.7$, which is close to the critical exponent of the two-dimensional Anderson model with SOC \cite{Queiroz2015, Asada2002}. This suggests that, in the strong-SOC regime, the quantum percolation transition is governed by the same universality class as the Anderson transition in two-dimensional disordered systems with time-reversal symmetry and broken SU(2) spin-rotation symmetry.


\section{Conclusions}
\label{conclusions}

In this work, we studied quantum percolation in a honeycomb lattice in the presence of random spin--orbit coupling. By combining exact diagonalization, level-statistics analysis, and finite-size scaling of the participation-entropy approximant $D_2$, we characterized the metal--insulator transition and its critical properties. For the case without spin--orbit coupling, we obtained a quantum percolation threshold $p_q=0.91(1)$, which lies significantly above the classical site-percolation threshold of the honeycomb lattice. This confirms that geometrical connectivity alone is not sufficient to guarantee extended quantum states, since quantum interference still suppresses transport over a substantial part of the percolating regime.

When the spin-orbit coupling is present, the first clear effect appears in the spectral correlations. The average ratio of adjacent gaps evolves from the Gaussian orthogonal ensemble value toward the Gaussian symplectic ensemble value as $\mu/t$ increases, signaling the crossover from the orthogonal to the symplectic symmetry class. This crossover is accompanied by a substantial modification of the metal--insulator transition itself. In particular, the quantum percolation threshold decreases monotonically as the spin–orbit coupling increases, indicating that spin–orbit coupling promotes delocalization and enables extended states to survive down to lower occupation probabilities. For larger spin-orbit coupling, the threshold typically saturates at about $p_q \approx 0.84$, which is still higher than the classical percolation threshold. Thus, even in the symplectic regime, quantum interference remains relevant and prevents the quantum transition from collapsing onto the purely geometrical one. At the same time, the critical exponent  $\nu$ decreases as the spin–orbit coupling is increased and tends toward $\nu \approx 2.7$, which is close to the established value for the two-dimensional symplectic universality class. This provides strong evidence that, in the regime of strong spin-orbit coupling, the transition in the diluted honeycomb lattice is controlled by the same universal critical behavior as the Anderson transition in two-dimensional disordered systems that preserve time-reversal symmetry but lack spin-rotation invariance.


From an experimental standpoint, these findings are directly pertinent to graphene-based platforms in which both disorder and spin-orbit coupling can be engineered. For instance, proximity-induced SOC in graphene or in transition-metal dichalcogenides (such as WS$_{2}$ or WSe$_{2}$) \cite{Wang2015,Wakamura2019,Yang2017}, as well as the deposition of adatoms (e.g., hydrogen, fluorine, or heavy elements), offer practical means to amplify the spin-orbit coupling effects while simultaneously introducing disorder. In addition, defect engineering via irradiation or chemical functionalization enables control over the vacancy concentration, effectively implementing site dilution \cite{Basta2023}. In such systems, the combined influence of disorder and SOC can be investigated through transport measurements \cite{Wang2015, Avsar2014}. In particular, magnetotransport experiments exhibit a crossover from weak localization to weak antilocalization  \cite{Wakamura2021}, providing a direct hallmark of the underlying symmetry-class transition.



\color{black}

\section*{ACKNOWLEDGMENTS}
J.F. acknowledges the support from ANID Fondecyt grant number 3240320. W.A.M.M acknowledges CNPq, Brazil (Grant No. 308560/2022-1). Powered@NLHPC: This research was partially supported by the NLHPC supercomputing infrastructure (CCSS210001).



\bibliography{qp-honey}
\end{document}